\documentclass[twocolumn,floatfix,showpacs,prd,aps,tightenlines,superscriptaddress]{revtex4}
\usepackage{graphicx}
\usepackage{psfrag}
\usepackage{dcolumn}
\usepackage{bm}

\newcommand{\bea}{\begin{eqnarray}}
\newcommand{\eea}{\end{eqnarray}}
\newcommand{\beq}{\begin{equation}}
\newcommand{\eeq}{\end{equation}}

\begin{document}

\title{Spin Flips and Precession in Black-Hole-Binary Mergers}

\author{Manuela Campanelli}
\affiliation{Center for Computational Relativity and Gravitation,
School of Mathematical Sciences,
Rochester Institute of Technology, 78 Lomb Memorial Drive, Rochester,
 New York 14623}
\affiliation{Center for Gravitational Wave Astronomy, Department of Physics and Astronomy,
The University of Texas at Brownsville, Brownsville, Texas 78520}

\author{Carlos O. Lousto} \affiliation{Center for Gravitational Wave Astronomy, Department of Physics and Astronomy,
The University of Texas at Brownsville, Brownsville, Texas 78520}
\affiliation{Center for Computational Relativity and Gravitation,
School of Mathematical Sciences,
Rochester Institute of Technology, 78 Lomb Memorial Drive, Rochester,
 New York 14623}

\author{Yosef Zlochower} \affiliation{Center for Gravitational Wave Astronomy,
Department of Physics and Astronomy, The University of Texas at Brownsville, Brownsville, Texas 78520}
\affiliation{Center for Computational Relativity and Gravitation,
School of Mathematical Sciences,
Rochester Institute of Technology, 78 Lomb Memorial Drive, Rochester,
 New York 14623}

\author{Badri Krishnan}
\affiliation{Max-Planck-Institut f\"ur Gravitationsphysik,
Albert-Einstein-Institut, Am M\"uhlenberg 1, D-14476 Golm, Germany}

\author {David Merritt} 
\affiliation{Department of Physics, 85 Lomb Memorial Drive, 
Rochester Institute of Technology, Rochester, NY 14623}
\affiliation{Center for Computational Relativity and Gravitation,
School of Mathematical Sciences,
Rochester Institute of Technology, 78 Lomb Memorial Drive, Rochester,
 New York 14623}

\date{\today}

\begin{abstract}
We use the `moving puncture' approach to perform fully non-linear
evolutions of spinning quasi-circular black-hole binaries with individual spins 
unaligned with the orbital angular momentum. 
We evolve configurations with the individual spins (parallel and equal in
magnitude)  pointing in the orbital plane and $45^\circ$ above the
 orbital plane.
We introduce a technique to measure the spin direction and track the
precession of the spin during the merger, as well as measure the spin
flip in the remnant horizon.
The former configuration completes
1.75 orbits before merging, with the spin precessing by $98^\circ$
 and the final remnant
horizon spin flipped by $\sim72^\circ$ with respect to the component
spins. The latter configuration completes 2.25 orbits, with the spins 
precessing by $151^\circ$
and the final remnant horizon spin flipped $\sim34^\circ$
with respect to the component spins.
These simulations show for the first time how the spins are reoriented during 
the final stage of black-hole-binary mergers verifying the hypothesis of the 
spin-flip phenomenon.
We also compute the track of the holes before merger and observe a
precession of the orbital plane with frequency similar to the orbital
frequency and amplitude increasing with time.
\end{abstract}

 \maketitle

\section{Introduction}

There is widespread interest in understanding the dynamics of the
last orbital stages of generic black-hole binaries (i.e.\ binaries
with randomly aligned spins and unequal masses). These last few
orbits before the binaries merge involve highly-nonlinear
gravitational interactions leading not only to large amounts of
gravitational radiation leaving the systems, but also to intricate
coupling effects, particularly those involving the spins of the two
component black holes. Questions of how the black holes orient their spins with
respect to the (instantaneous) orbital plane and the extent to
which the black holes spin-up during the
last orbital stages are of great astrophysical interest. Spinning
black holes are believed to be the engines of active
galactic nuclei and quasars; the efficiency with which infalling
matter is converted into radiation depends on the energy of the
innermost stable orbit and is greatest for rapidly-spinning
holes~\cite{1999agnc.book.....K,2002ApJ...565L..75E}.
The energetic jets observed in many AGN and stellar-mass accreting
systems are believed to be launched perpendicularly to the inner
accretion disk, hence parallel to the spin axis of the accreting
hole~\cite{1984RvMP...56..255B}.
Changes in the spin direction are therefore potentially observable.
A number of active galaxies exhibit semi-periodic deviations of
the jet directions from a straight line~\cite{2006MmSAI..77..733K}, suggestive
of precession of the accretion disk around the jet-emitting hole
or geodetic precession of the larger hole, either of which might be
driven by torques from a second orbiting black
hole~\cite{1993ApJ...409..130R,1997ApJ...478..527K}.
About 15 radio galaxies show jets with apparently abrupt changes in
jet direction, forming X-shaped
patterns~\cite{1985A&AS...59..511P,1992ersf.meet..307L}.
Jets of Seyfert galaxies often misalign from subkiloparsec
to kiloparsec scales~\cite{2006AJ....132..546G}.
Some models~\cite{Merritt:2002hc,2002A&A...396...91Z}  attribute
these spatial variations to a sudden re-orientation of the spin axis
of the larger hole as it accretes a smaller hole.
Fossil evidence that such re-orientations were common in the past
is observed in the nearly random orientations of jets in disk galaxies
with respect to the disk
plane~\cite{2000ApJ...537..152K,2002ApJ...568..998M}.

In the generic case, inspiral of a black-hole binary should induce
a precession in the direction of the spin axis of
either hole~\cite{2005LRR.....8....8M}, and unless the mass ratio is
extreme, the final spin orientation is dominated by the orbital
angular momentum of the binary, implying substantial re-orientation as
long as the spin and orbital angular momenta are initially
mis-aligned~\cite{Merritt:2002hc}. We refer to this jump in the spin
direction of the remnant, with respect to the individual spins, as a
spin-flip. Note that this definition of spin-flip does not require
non-linear interactions; it is a simple consequence of the remnant
acquiring most of the total angular momentum of the original binary.
Also note that the `flip' is actually the difference in spins between
two distinct types of objects: individual horizons and common horizons.

Aside from their astrophysical interest, black-hole binaries are one
of the primary targets of the earth-based gravitational wave
observatories, such as LIGO~\cite{LIGO}, VIRGO~\cite{VIRGO},
GEO600~\cite{geo_web}, etc.  These detectors are operating at
unprecedented sensitivities. In particular, the LIGO interferometers
are currently taking data at their design sensitivities.  The initial
LIGO interferometers might be able to observe two $20 M_\odot$
inspiraling black holes out to a distance of more than 100Mpc, and
advanced LIGO could see the same event up to a cosmological redshift of
$z\approx 0.4$~\cite{Cutler:2002me}.  Predicted black-hole-binary
gravitational waveforms will not only be of great help for the
detection of this radiation using matched filtering techniques, but
will also be essential for the interpretation of the signals,
determination of the event rates, and extraction of astrophysical
parameters. This will be especially important for the next generation
of ground and space-based detectors, such as LISA~\cite{LISA,
  Danzmann:2003tv} which should observe gravitational wave bursts from
the mergers of supermassive black holes in the centers of galaxies out
to very high redshifts.

Simulations of the last orbital stages of black-hole binaries require
solving the fully-nonlinear General Relativity field equations numerically on 
supercomputers. However, solving these non-linear equations proved to
be quite difficult and the problem remained unsolved for over thirty
years. However, last year two independent techniques were developed
that broke through the barrier of the numerical instabilities to produce
spectacular results. In
2005 these two approaches were used to generate the gravitational 
waveforms from the last orbit of
non-spinning, equal-mass black-hole binaries. The first technique,
which was developed by Pretorius~\cite{Pretorius:2005gq},
used a second-order formulation of the General
Relativity field equations in a generalized harmonic gauge (GHG), along with
singularity excision in the interior of
the horizons, adaptive mesh refinement (AMR) on a compactified space,
 and the addition of constraint damping terms to the evolution
equation.
The second successful approach was
developed a few months later by our group at UTB~\cite{Campanelli:2005dd}
and independently by the Numerical Relativity group at
NASA/Goddard~\cite{Baker:2005vv}. This latter method uses
a mixed first-and-second-order formulation of the General Relativity field
equations known
as BSSN system~\cite{Nakamura87,Shibata95, Baumgarte99}, in
combination with the puncture formalism~\cite{Brandt97b,
Brugmann:1997uc}
(without the need for singularity excision).
This technique differs from previous work with punctures in that the
punctures are not fixed on the grid.
In both versions of this `moving puncture' approach modifications to
the standard 1+log lapse and Gamma-driver shift conditions were
introduced~\cite{Alcubierre02a,Campanelli:2005dd,Baker:2005vv}.

Most of the groups in numerical relativity have now implemented one of
the two approaches; the `moving puncture' technique being the more
popular.  Given its technical simplicity and flexibility, this latter
approach has been used to  produce
several interesting results ranging from the original simulations
of non-spinning equal-mass
binaries~\cite{Campanelli:2006gf,Pretorius:2006tp,Baker:2006yw,Baker:2006ha} to spinning
equal-mass binaries~\cite{Campanelli:2006uy,Campanelli:2006fg} and
non-spinning unequal-mass binaries~\cite{Herrmann:2006ks,Baker:2006vn,Gonzalez:2006md}. Notably, the
moving puncture technique has recently also been successfully
implemented in the case of
neutron-star---black-hole-binaries~\cite{Shibata:2006bs,Loffler:2006nu},
thus extending the impact of the technique on the
numerical/astrophysical relativity community. Noteworthy, the GHG approach
has also been successfully used to study eccentric orbits
in black-hole binaries with equal-mass and small corotation spins~\cite{Pretorius:2006tp}.
A generalized harmonic form of Einstein's equations has 
also been successfully used together with a dual coordinates method 
to evolve black-hole-binary spacetimes for several orbits prior
merger~\cite{Scheel:2006gg}.

Alongside this impressive progress in computational and experimental
relativity, the last few years have also seen significant mathematical
progress in our understanding of black holes in full non-linear
General Relativity.  There is now a better understanding of the
geometry and dynamics of trapped and marginally trapped surfaces using
the quasi-local notions of trapping~\cite{Hayward:1993wb}, isolated
\cite{Ashtekar:2000sz} and dynamical horizons~\cite{Ashtekar:2002ag}.
Isolated horizons describe black holes in equilibrium in
non-stationary spacetimes, and trapping and dynamical horizons
describe the general time-dependent case.  The applications of these
ideas to classical and quantum black hole physics are too numerous to
be described here, and we refer to
\cite{Ashtekar:2004cn,Gourgoulhon:2005ng,Booth:2005qc} for reviews and
a more complete set of references.  For our present purposes, we are
mostly interested in calculating the angular momentum of an
(approximately) axisymmetric horizon.  The calculation of angular
momentum for isolated horizons is carried out using Hamiltonian
methods as described in
\cite{Ashtekar:2000hw,Ashtekar:2001is,Booth:2001gx}. The analogous
Hamiltonian calculation for non-stationary trapping and dynamical
horizons is given in~\cite{Booth:2005ss}.  Conservation and balance
laws describing how the horizon mass and angular momentum change in
response to infalling matter/radiation are found in
\cite{Ashtekar:2003hk,Hayward:2004dv,Hayward:2006ss,Gourgoulhon:2006uc}.
The calculation of the magnitude of black hole spin angular momentum
tailored to numerical relativity is presented in~\cite{Dreyer02a} and
more recently, \cite{Schnetter:2006yt} considers higher mass and
angular momentum multipole moments.  
In this paper we use this formalism primarily to
compute the direction and magnitude of the spin angular momentum
vector.

This paper is organized as follows. In Sec.~\ref{sec:pn} we discuss
the Post-Newtonian predictions for spin and orbital plane precession.
In Sec.~\ref{sec:techniques} we
describe the techniques used to evolve the binary and measure the
horizon spins. In Sec.~\ref{sec:id} we describe the initial
data parameters for the binary configurations mentioned in the
remainder of the text. 
In Sec.~\ref{sec:results} we give a detailed description of the new results
regarding spin and orbit precession. In Sec.~\ref{sec:summary} we
discuss some of the implications of our results. Finally, in
Appendix~\ref{sec:review} we review past results from aligned-spin
and non-spinning binaries.

\section {Post-Newtonian analysis}
\label{sec:pn}
If the spins of the two black holes in a binary are not aligned with the total 
angular momentum, then the spin and orbital angular momenta will 
precess about the total angular momentum. 
Precession of the spin of the holes 
is produced by spin-orbit coupling and the spin-spin coupling. This effect 
has been studied in several papers by means of the Post-Newtonian 
expansion~\cite{Barker79b,Kidder:1995zr} 
\begin{eqnarray}\label{sidot} 
\frac{d\vec S_i}{dt} = {1 \over r^3} \biggl\{ && ({\vec L_N \times \vec S_i})(2+{3 \over 2} 
{m_j \over m_i}) - {\vec S_j \times \vec S_i} \nonumber \\ && \mbox{} 
+ 3({ \hat n \cdot \vec S_j}) 
{\hat n \times \vec S_i} \biggr\} , 
\end{eqnarray} 
where ${\vec L_N} \equiv \mu ({\vec x \times \vec v})$ is the Newtonian orbital 
angular momentum,
${\vec x} \equiv {\vec x_1}-{\vec x_2}$, $r \equiv |\vec x|$,
${\vec v}={d{\vec x}/dt}$, ${\hat n}\equiv{{\vec x}/r}$, $m=m_1+m_2$,
$\mu \equiv m_1m_2/m$, $\eta \equiv \mu /m$,
${\vec S} \equiv {\vec S_1}+{\vec S_2}$, an over-dot 
denotes $d/dt$, and the $j$ subscript denotes the other hole.
Since the time derivative of the individual spin of the holes has the 
form $\dot{\vec {S}}_i=\vec{\Omega}_i\times\vec{S}_i$ this implies that the 
magnitude of each individual spin is conserved and only its direction will
continuously change in time and rotate at a precession
frequency given by 
\begin{eqnarray} 
\vec{\Omega}_i={1 \over r^3} \biggl\{ (2+{3 \over 2} 
{m_j \over m_i})\,\vec{L}_N - \vec{S}_j 
+ 3({\hat n \cdot \vec{S}_j}){\hat n} \biggr\}.
\end{eqnarray} 

Both the spin and orbital planes precess. This is due to the fact
that, if the radiated angular momentum is neglected, the total angular
momentum $\vec J = \vec L + \vec S$ is conserved. Hence the $\dot
{\vec L} = - \dot {\vec S}$.  Thus the orbital plane will precess at
the same frequency as the total spin. The gravitational radiation
reaction effect will generate a net loss of $J$ that will actually
produce an increase in the amplitude of the spin and orbital
oscillations since the black holes will get closer; magnifying the
spin-orbit coupling.  In this paper, we go beyond the Post-Newtonian
expansion and study these effects using full numerical evolutions.

Precession occurs in the plane perpendicular to the total angular
momentum. We thus split the spin vector of each hole into components
parallel to, and perpendicular to, the total angular momentum, i.e.\
$\vec S = \vec S_\parallel + \vec S_\perp$, where $\vec S_\parallel =
(\vec S \cdot \hat J) \hat J$ and $\vec S_\perp = \vec S - (\vec
S\cdot \hat J) \hat J$.
Note that the direction (but not magnitude) of the total angular
momentum does not change significantly between the start and end of the
simulations. We define the total angle of precession $\Theta_p$ as the
angle by which $\vec S_\perp$ is rotated between the start and the end of
the simulation. The total precession angle is then given by
\begin{equation}
\label{eq:precessangle}
  \cos \Theta_p = \frac{\vec S_M \cdot \vec S_I -
     (\hat J\cdot \vec S_M)(\hat J\cdot \vec S_I)}
     {\sqrt{[S_M^2 - (\hat J\cdot\vec S_M)^2]
     [S_I^2 - (\hat J\cdot\vec S_I)^2]}},
\end{equation}
where $\vec S_I$ is the initial spin of one of the individual horizons,
$\vec S_M$ is the spin of that individual horizon at the merger time,
and $\vec J$ is the initial total angular momentum of the system.

The relevance of precessing spinning-black-hole binaries to data
analysis has been stressed in several papers using the post-Newtonian
templates. `Spiky' templates have been considered to detect moderate
massive galactic binaries
in~\cite{Grandclement:2002dv,Grandclement:2002vx,Grandclement:2003ck}.
Detection and post-Newtonian dynamics in precessing binaries have also been
extensively discussed
in~\cite{Buonanno:2002fy,Pan:2003qt,Buonanno:2004yd,Hartl:2004xr,Buonanno:2005xu,Buonanno:2005pt},
and the relevance to LISA observations has been discussed
in~\cite{Vecchio:2003tn}.

\section{Techniques}
\label{sec:techniques}

We use the Brandt-Br\"ugmann puncture approach~\cite{Brandt97b} along
with the {\sc TwoPunctures}~\cite{Ansorg:2004ds} and {\sc
BAM\_Elliptic}~\cite{cactus_web} thorns to compute initial data.  In
this approach the 3-metric on the initial slice has the form
$\gamma_{a b} = (\psi_{BL} + u)^4 \delta_{a b}$, where $\psi_{BL}$ is
the Brill-Lindquist conformal factor, $\delta_{ab}$ is the Euclidean
metric, and $u$ is (at least) $C^2$ on the punctures.  The
Brill-Lindquist conformal factor is given by
$
\psi_{BL} = 1 + \sum_{i=1}^n m_i / (2 r_i),
$
where $n$ is the total number of `punctures', $m_i$ is the mass
parameter of puncture $i$ ($m_i$ is {\em not} the horizon mass
associated with puncture $i$), and $r_i$ is the coordinate distance to
puncture $i$. In all cases below, we evolve data containing only two
punctures with equal puncture mass parameters, and we denote this
puncture mass parameter by $m_p$.  We evolve these black-hole-binary
data-sets using the {\sc LazEv}~\cite{Zlochower:2005bj} implementation
of the moving puncture approach~\cite{Campanelli:2005dd,
Baker:2005vv}.  In our version of the moving puncture
approach~\cite{Campanelli:2005dd} we replace the
BSSN~\cite{Nakamura87,Shibata95, Baumgarte99} conformal exponent
$\phi$, which has logarithmic singularities at the punctures, with the
initially $C^4$ field $\chi = \exp(-4\phi)$.  This new variable, along
with the other BSSN variables, will remain finite provided that one
uses a suitable choice for the gauge. An alternative approach uses
standard finite differencing of $\phi$~\cite{Baker:2005vv}. Note that
both approaches have been used successfully by several other
groups~\cite{Sperhake:2006cy, Herrmann:2006ks,
Hannam:2006vv,Gonzalez:2006md,Bruegmann:2006at,Shibata:2006bs}.

We obtain accurate, convergent waveforms and horizon parameters by
evolving this system in conjunction with a modified 1+log lapse, a
modified Gamma-driver shift
condition~\cite{Alcubierre02a,Campanelli:2005dd}, and an initial lapse
$\alpha\sim\psi_{BL}^{-4}$.  The lapse and shift are evolved with
$(\partial_t - \beta^i \partial_i) \alpha = - 2 \alpha K$, $\partial_t
\beta^a = B^a$, and $\partial_t B^a = 3/4 \partial_t \tilde \Gamma^a -
\eta B^a$.
These gauge conditions require careful treatment of $\chi$
near the puncture in order for the system to remain
stable~\cite{Campanelli:2005dd,Campanelli:2006gf,Bruegmann:2006at}. In
Ref.~\cite{Gundlach:2006tw} it was
shown that this choice of gauge leads to a strongly hyperbolic
evolution system provided that the shift does not become too large.
For our version of the moving puncture approach, we find that the
product $\alpha \tilde A^{ij} \partial_j \phi$ initially has to be
$C^4$ on the puncture. In the spinning case, $\tilde A^{ij}$ is
$O(r^3)$ on the puncture, thus requiring that $\alpha \propto r^3$ to
maintain differentiability. We therefore choose an initial lapse
$\alpha(t=0) = 2/(1+\psi_{BL}^{4})$ which is $O(r^4)$ and $C^4$ on the
puncture and reproduces the isotropic Schwarzschild lapse at large
distances from the horizons.  The initial values of $\beta^i$ and
$B^i$ are set to zero.

Hannam~et.~al.~\cite{Hannam:2006vv} examine the smoothness of the
evolved fields at late
times at the puncture. They find that, in the case of Schwarzschild,
 $\chi$ transitions from an initially $C^4$
field to a $C^2$ field at late times. Although we require that the
fields be
initially $C^4$, this late-time drop in smoothness does not
appear to leak out of the horizon (which is consistent with the
analysis in~\cite{Hannam:2006vv}).

We use a `multiple transition' fisheye
transformation~\cite{Campanelli:2006gf} to push the boundaries to
$200M$, while maintaining a resolution of up to $M/30$ in the central
region.

We measure the magnitude $S$ of the angular momentum of the horizons
using our implementation of the algorithm detailed in~\cite{Dreyer02a}. The
magnitude of the horizon spin is given by
\begin{equation}\label{isolatedspin} 
S = \frac{1}{8\pi}\oint_{AH}(\varphi^aR^bK_{ab})d^2V 
\end{equation} 
where $\varphi^a$ is an approximate Killing vector on the horizon,
$K_{ab}$ is the extrinsic curvature of the 3D-slice, $d^2V$ is the
natural volume element intrinsic to the horizon, and $R^a$ is the
outward pointing unit vector normal to the horizon on the 3D-slice;
the sign of $\varphi^a$ is chosen so that $S$ is positive.  This
algorithm for calculating $S$ was initially meant to be applied to the
case when the individual black holes are modeled as axisymmetric
isolated horizons, which is valid when the two black holes are
sufficiently far away from each other.  The isolated horizon formalism
is generalized to the dynamical case through the notion of a dynamical
horizon~\cite{Ashtekar:2003hk}, and the formula for $S$ remains valid
under this generalization.

Turning now to the direction of the spin angular momentum vector, we
first note that, in general, it seems difficult to assign a unique
coordinate independent 3-vector $\vec{S}$ to a spinning horizon.  For
example, we could take a normal Kerr spacetime and slice it
non-axisymmetrically so that it becomes difficult to assign a spin
3-vector $\vec{S}$ to the black hole on these distorted 3D-slices.
There is however a generalization which works.  To see this, first
note that \emph{every} smooth cross section (with complete $S^2$
topology) of a Kerr horizon is axisymmetric, no matter how distorted
this cross-section is.  This may seem somewhat surprising at first
glance, but it is a straightforward consequence of the fact that the
null generators of the Kerr horizon have vanishing expansion, shear
and twist; the axial symmetry vector projects to a symmetry of the
2-geometry of the cross-section.  Thus there exists a symmetry vector
$\varphi^a$ on this cross-section.  The poles of the horizon are then
defined to be the points where the axial symmetry vector $\varphi^a$
vanishes.  From a spacetime perspective, the locus of points on the
Kerr horizon defined by $\varphi^a\varphi_a = 0$ is a coordinate and gauge
independent notion.  These considerations remain valid on every
axisymmetric isolated horizon.  As long as we have a suitable axial
vector, we can similarly define the poles even for dynamical and
trapping horizons.  The poles exist whenever we can assign a (possibly
approximate) axial symmetry vector $\varphi^a$ on the horizon. Of
course, when the horizons become extremely distorted, it might happen
that it is no longer approximately axisymmetric, or the axial vector
might have more than two poles.  In such extreme cases, this would not
work.  But we shall see through our numerical simulations that there
is a significant dynamical regime where exactly 2 poles exist, and
these problems do not arise.

Given the location of the two poles on the horizon, how do we assign a
3-vector to them, and thereby obtain all the components of the spin
vector $\vec{S}$?  An obvious starting point would be to use the unit
normal vector $R^a$ at the poles.  This would not give a unique answer
in the absence of reflection symmetry.  Alternatively, we could
consider the curl of $\varphi^a$ suitably averaged over the horizon.
However, even if we could successfully assign such a 3-vector
uniquely, it is not clear in general how this vector should be
compared with the spin 3-vector calculated at spatial infinity.  This
could be done in spacetimes with global axisymmetry, but this is not
available to us in the present case.  In the absence of a solution to
this problem, we simply define the direction of the spin to be the
Euclidean unit-norm vector tangent to the coordinate line joining the
two poles. The spin-vector $\vec S_{\rm IH}$ is then equal to this
Euclidean unit-norm vector multiplied by the Isolated Horizon spin
obtained from Eq.~(\ref{isolatedspin}).  The definition of $\vec{S}$
might need to be further refined, however it seems to be satisfactory
for our purposes. This definition of the spin vector reproduces the
Bowen-York spin parameters on the initial slice, and should remain
reasonable as long as the coordinates do not become too distorted.  In
addition to using the Killing vector $\varphi^a$, we also found it
useful to define angular momenta with the flat space coordinate
rotational killing vectors
\begin{eqnarray*}
 \varphi^a_x &=& \left[0, -(z-z_c), (y-y_c)\right],\\
 \varphi^a_y &=& \left[(z-z_c), 0, -(x-x_c)\right],\\
 \varphi^a_z &=& \left[-(y-y_c), (x-x_c), 0\right],
\end{eqnarray*}
where $(x_c, y_c, z_c)$ is the coordinate centroid of the horizon.
 We can then obtain the
coordinate-base spin vector $\vec S_{\rm coord} = (S_x, S_y, S_z)$ by
replacing the approximate Killing vector in Eq.~(\ref{isolatedspin})
with the three coordinate rotational vectors
 (i.e.\ $S_i =\frac{1}{8\pi}\oint_{AH}
(\varphi_i^aR^bK_{ab})\,\textrm{d}^2V $). This definition of the spin
direction reproduces the Bowen-York spin parameters on the initial
slice as well, and produces reasonable results at later times for the
gauges used here.  (Of course this latter coordinate based calculation
will not yield an accurate evaluation of the spin direction or
magnitude for more general gauges, while the former approximate
Killing vector calculation will produce accurate spin magnitudes for
generic gauges.)  In both cases the spin magnitude is the Euclidean
norm of the 3-vector $\vec S$. Note that in the former case this
Euclidean norm is precisely the spin given by Eq.~(\ref{isolatedspin}).

We solve for the approximate Killing vector field $\varphi^a$ on the
horizon using standard spherical-polar coordinates. In these
coordinates the Killing vector is obtained with highest accuracy when
its poles are aligned with the coordinate poles of the $(\theta,
\phi)$ coordinates. To make the calculation as accurate as possible,
we find the minimum of $\varphi^a\varphi_a$ in the northern hemisphere
and then rotate the angular coordinates so that the new north pole is
aligned with the minimum of $\varphi^a\varphi_a$. We then re-calculate
$\varphi^a$ to obtain a more accurate location of the minimum and
iterate until the new minimum of $\varphi^a\varphi_a$ lies on the
north pole.  There is a complication in the above procedure in that we
cannot calculate $\varphi^a$ on the coordinate poles themselves
(since the 2D Christoffel symbol is singular).  In
practice we stop iterating when the minimum of $\varphi^a\varphi_a$
lies within 2 angular grid-points of the coordinate pole. The
spin-direction associated with the minimum of $\varphi^a\varphi_a$
therefore cannot be obtained with higher precision than a few angular
grid sizes.  It might be possible to improve the accuracy by
considering multiple patches on the horizon to avoid the coordinate
singularity, or to use a spectral decomposition.

We found that using $160$ points in the $\theta$ direction and $320$
points in the $\phi$ direction provides reasonable results for the
horizon spin calculation, with errors in the spin direction of about
$2^\circ$. Adding significantly more points only increases
the numerical error because the horizon algorithm itself uses far
fewer points to locate the apparent horizons and the underlying
numerical grid has a far coarser resolution.

The configurations discussed in this paper contain either PI-symmetry,
i.e.\ $(x,y,z) \to (-x,-y,z)$, or parity-symmetry, i.e.\
$(x,y,z)\to(-x,-y,-z)$. We exploit these two symmetries to reduce the
grid size by a factor of 2. The zero-spin and (anti-)aligned
spin binaries have the additional symmetry $(x,y,z)\to(x,y,-z)$.
We implement the parity-symmetry boundary conditions using a locally
modified version of the PI-symmetry boundary thorn kindly provided to
us by Erik Schnetter. Note that behavior of the components of $g_{ab}$
and $k_{ab}$ under these symmetries can be obtained from the behavior of
these components under simple reflections. Thus, $g_{xz} (x,y,z) = g_{xz}
(-x,-y,-z)$ under parity-symmetry while $g_{xz}(x,y,z) = - g_{x
z}(-x,-y,z)$ under PI-symmetry. The Einstein equations preserve these
symmetries. In the case of spinning, equal-mass binaries,
parity-symmetry requires that the two spins be equal in magnitude and
parallel (i.e.\ the spin-vector behaves as a pseudo vector).

\section {Initial Configurations}
\label{sec:id}
We study two configurations of non-aligned-spin binaries with
parallel spins (equal in magnitude) that exhibit
spin and orbital-plane precession, as well as spin-flips of the remnant
horizon spin with respect to the individual horizon spins. We choose
 configurations
where the binary separation is small enough that the spin-orbit
coupling is
large, but large enough that the binaries complete at least $\sim1.75$
orbits before merging. The first configuration,
which we denote with SP3 starts with the spins aligned along the
initial orbital plane. This can be interpreted
as a binary in which one black hole orbits about the pole
of the second black hole.  The second configuration, which we denote
with SP4, starts with
the spins pointing $45^\circ$ above the initial orbital plane,
corresponding to a binary in which infall
occurs initially along a plane tilted with respect to both
spins.  In
both cases the masses and spins of the two holes are equal (i.e.\
spins parallel and equal in magnitude).
Setting the two masses and spins equal ensures that the system
is parity-symmetric,
but still is generic enough to display both spin and orbital plane precession
as well as a spin-flip in the direction of the orbital angular momentum.
The initial data parameters for these two
configurations, which were obtained using the 3PN equations
of motion, are given in Table~\ref{table:ID}. We also report the
initial-data parameters for the previously studied aligned (S++, $SC$),
anti-aligned (S-\,-), and non-spinning binaries ($S0$).
The PN data provides the puncture location, momenta, and spins. We complete
the data by choosing puncture mass parameters (equal for the two
punctures) such that the total ADM mass of the system is 1.

\begin{widetext}

\begin{table}
\caption{Initial data for quasi-circular, equal-mass black-hole
binaries.
The binaries have an ADM mass of $(1.0000\pm0.0005)M$, with
orbital frequency $M\Omega$ fixed to $0.0500$, and initial
proper separations $l$.
 The punctures
are located at $(\pm X, 0,0)$, with mass parameter $m_p$, momentum
$(0,\pm P,0)$, spin angular momentum $(0,S_y,S_z)$, and specific spin
$S/m^2$ ($m$ is the horizon mass).
}
\begin{ruledtabular}
\begin{tabular}{lllllllll}\label{table:ID}
Name & $S_y/M^2$ & $S_z/M^2$ & $X/M$ & $P/M$ & $J/M^2$ & $S/m^2$ & $l/M$
&$m_p/M$\\
\hline
SP3 & $0.128725$ & $0$ & $3.276347$ & $0.133587$ &
$0.91243$ &$0.5013$ & $10.20$ & $0.43025$ \\
SP4 & $0.091198$ & $0.091198$ & $3.179908$ & $0.1314406$ &
$1.03454$ & $0.5007$ & $9.94$ & $0.43037$\\
$S0$ & 0.0 & 0.0 & 3.280 & 0.1336 & 0.876 & 0.0 & 10.01 & 0.4848\\
$SC$ & 0.0 & 0.025757 & 3.2534 & 0.1330 & 0.917 & 0.1001 & 9.93 &
0.4831\\
S++ & 0.0 & 0.1939 & 3.0595 & 0.1291 & 1.1778& 0.757 & 9.27 & 0.3344\\
S-\,- & 0.0 & -0.1924 & 3.465 & 0.1382 & 0.5729 & -0.757 & 10.3 &0.3344 \\
\end{tabular}
\end{ruledtabular}
\end{table}

\end{widetext}

\section{Results}
\label{sec:results}
We evolved the SP3 configuration using central resolutions of
$h=M/22.5$, $h=M/25$, and $h=M/30$; with grid-sizes of 
$576^2\times288$, $640^2\times320$, and
$768^2\times384$ respectively. We used `multiple transition' fisheye
transformation~\cite{Campanelli:2006gf}
to place the outer boundaries at $200M$; far enough away that
boundary effects do not interfere with the orbital dynamics of the
system. In addition, we also evolved the SP4 configuration with a
central resolution of $h=M/25$, a grid size of
$640^2\times320$, and outer boundary at $200M$.

\begin{figure}
\begin{center}
\includegraphics[width=3.3in]{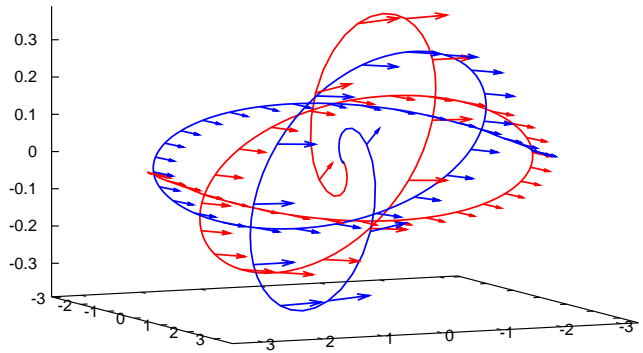}
\caption{The puncture trajectories along with spin direction
(every $4M$) for the SP3 configuration
for the $M/30$ resolution run. The spins are initially aligned along
the $y$-axis,  but rotate by $\sim 90^\circ$ during
the 1.25 last orbits and also acquire a non-negligible $z$-component.
Note that the $z$-scale is 1/10th the x and y scale.}
\label{fig:sp3_track_spin}
\end{center}
\end{figure}

\begin{figure}
\begin{center}
\includegraphics[width=3.3in]{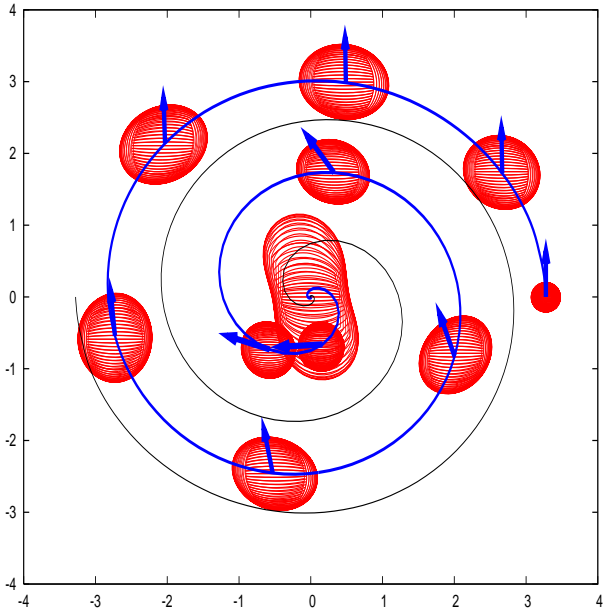}
\caption{The projection of puncture trajectories and spin for the SP3 configuration
onto the $xy$ plane along with the individual apparent horizons for
the $M/30$ run. The horizons and spins are shown at $t=0, 20M, 40M,
..., 160M, 164M$. The first common horizon (also shown) formed at
$t=164.2M$. The spins are initially aligned along the $y$-axis 
but rotate by $\sim 90^\circ$ during the last 1.25 orbits.
The spin of the second black hole (not shown) is equal to the first.
}
\label{fig:sp3_track_spin_xy}
\end{center}
\end{figure}

Figures~\ref{fig:sp3_track_spin}~and~\ref{fig:sp3_track_spin_xy}
show the puncture trajectory and horizon-spin direction along this
track for the SP3 configuration (the latter
suppressing the $z$-direction). Note that
the scale of the $z$-axis in Fig.~\ref{fig:sp3_track_spin} is 1/10th that
 of the $x$ and $y$ axes. From the
plots one can clearly see the orbital plane precess out of the equatorial
plane, as well as the spin axis rotating by approximately $90^\circ$ in the $xy$ plane
 during the course
of the merger. The spins are initially aligned along the $y$-axis,
 but at merger they show both a significant $z$-component
and an approximate $90^\circ$ rotation to the $-x$-axis.
The individual horizon spins at the merger are $\vec S_{\rm coord} =
(-0.121\pm0.002, -0.007\pm0.003, 0.037\pm0.003)$ (we use the
coordinate based measure of the spin at the merger because the
calculation of $S_{\rm IH}$ is not accurate when the black holes
are this close together; see comments below). Hence 
the total precession angle for the SP3 configuration is $\Theta_p = 98^\circ$.
Note that there is no discernible correlation between the
orientation of the projected horizon and the projected spin direction.

\begin{figure}
\begin{center}
\includegraphics[width=3.3in]{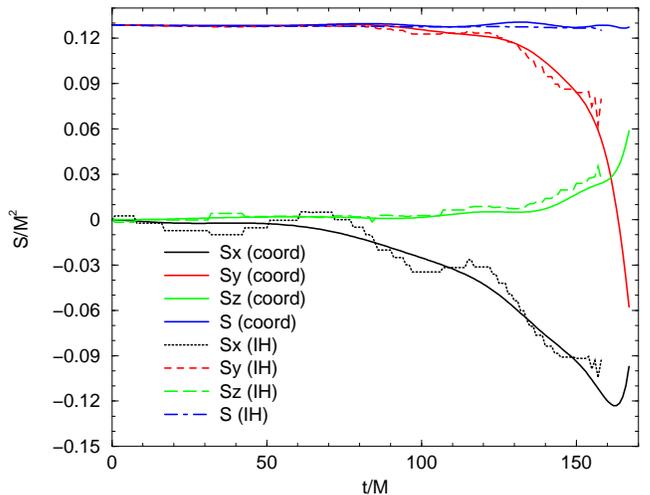}
\caption{The spin components and magnitude
 versus time for the SP3 configuration as
calculated using the coordinate rotational vectors (coord) and the poles of
the approximate Killing vector (IH) for the $M/30$ resolution.  The
calculation of the approximate Killing vector breaks down near the
merger (which occurs at $t=164.2$), but the purely coordinate based
calculation continues to produce reasonable results. Note that the
direction obtained from the Killing vector oscillates about the
coordinate based direction. Also note the spin has just rotated by
$90^\circ$ in the xy plane at the time of merger. The spin magnitude
remains essentially constant throughout the merger phase. The
magnitude of the spin calculated from the Killing vector is coordinate
invariant and, unlike the spin direction, is expected to be more
accurate than the coordinate based calculation.   }
\label{fig:sp3_spin_dir}
\end{center}
\end{figure}

\begin{figure}
\begin{center}
\includegraphics[width=3.3in]{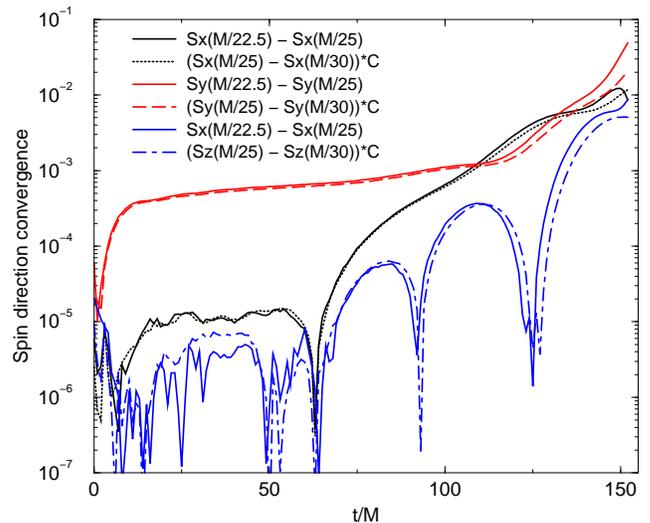}
\caption{A convergence plot of the coordinate-base $\vec S_{\rm
coord}$ calculation for the SP3 configuration. 
Note that for our choice of
resolutions, third-order convergence is demonstrated by
$S(M/22.5) - S(M/25) = (S(M/25) - S(M/30))  C$, where 
$C=0.88$. The spin is initially third-order convergent, with
higher order-convergence apparent at later times. }
\label{fig:sp3_spin_dir_conv}
\end{center}
\end{figure}

\begin{figure}
\begin{center}
\includegraphics[width=3.3in]{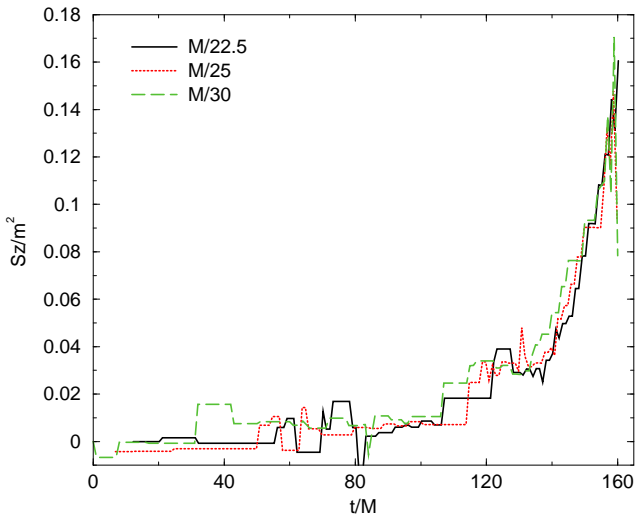}
\caption{The $z$-component of the Killing vector based calculation
of the spin rescaled by the square of the horizon mass for the three
resolutions. The curves have been translated by a distance equal to
the difference in merger times of the $M/22.5$ and $M/25$ runs with
the merger time of the $M/30$ run. In this configuration, unlike the
previously studied aligned-spin binary, the `spin-up' appears to
be large. However, in this case the `spin-up' is actually the rotation
of the nearly-constant spin-vector towards the $z$-axis, rather than
an increase in the spin magnitude.}
\label{fig:sp3_kill_z}
\end{center}
\end{figure}

In Fig.~\ref{fig:sp3_spin_dir} we show the coordinate ($\vec S_{\rm
coord}$) and Killing vector based ($\vec S_{\rm IH}$) calculation of
the spin components versus time. $\vec S_{\rm IH}$ displays a step-function-like behavior due
to the difficulty in finding the poles (i.e.\ the zeroes of $\varphi^a
\varphi_a$) in the Killing vector accurately. As
discussed above, the Killing vector calculation is most accurate when
its poles are located at the coordinate poles of the $(\theta,\phi)$
coordinates on the horizon. However, the difficulty in calculating the
Killing vector itself near the coordinate poles introduces an
uncertainty in the location of the Killing vector poles. We also stress
that the 
direction associated with the location of the two poles of the Killing
vector is coordinate dependent. From the figure we see that the
$x$-and-$y$-components of $\vec S_{\rm IH}$ oscillates about the much
more regular $x$ and $y$ components of $\vec S_{\rm coord}$, while
the $z$-component of $\vec S_{\rm IH}$  is consistently larger than the
$z$-component of $\vec S_{\rm coord}$. The calculation of $S_{\rm IH}$
(and hence $\vec S_{\rm IH}$) breaks down
prior to the merger when the horizons get too close (and hence the mutual 
tidal distortions destroy the approximate axial symmetry). $\vec
S_{\rm coord}$, however, continues to produce reasonable results
through the merger. Thus, it is $\vec S_{\rm coord}$ that shows the
 clear rotation of the spin from the
$y$-axis to the $x$-axis at the merger. 
Note that the uncertainties in the spin directions do not correspond to
uncertainties in the magnitudes of the spin. For the Killing vector
based calculation $\vec S_{\rm IH}$, it is the spin magnitude that
is determined with high accuracy. Figure~\ref{fig:sp3_spin_dir_conv}
shows between third and fourth-order convergence of the components of
$\vec S_{\rm coord}$ from the three
resolutions (the third-order error may be due to third-order errors
leaking out of the puncture as well as third-order errors from the 
horizon calculations), while Fig.~\ref{fig:sp3_kill_z} shows the value of the
$z$-component of the specific spin $S_z/m^2$ (where $m$ is the horizon
mass)
based on the $z$-component of $\vec S_{\rm IH}$  for the three resolutions.
In this latter
figure the curves have been translated. (A convergence plot of $\vec
S_{\rm IH}$ would not be meaningful
 because the size of the step discontinuities in $\vec S_{\rm IH}$ are
larger than the differences in the spin direction
with resolution.) For the $z$-component of the spin,
we expect that, given the lack of significant oscillations plaguing
the x and y components, $\vec S_{\rm IH}$ gives a
better measurement than the more highly coordinate-dependent 
$\vec S_{\rm coord}$. The spin-up of the $z$-component of the specific spin
by $0.16$ is ten times larger than the analogous spin-ups of about $0.01$
seen in the zero-spin (S0) and aligned-spin (SC) configurations (see~\cite{Campanelli:2006fg}).
However, as can be seen in Fig.~\ref{fig:sp3_spin_dir}, the spin magnitude
does not increase significantly. Thus this spin-up in the $z$-direction
is not equivalent to the spin-up observed in the case of the aligned-spin
and non-spinning binaries. In those cases the spin-up involved an
increase in the spin magnitude, while here it primarily involves a
rotation of the spin vector out of the $xy$ plane. This rotation of
the spin out of the $xy$ plane follows the post-Newtonian predictions
of Eq.~(\ref{sidot}). 

\begin{figure}
\begin{center}
\includegraphics[width=3.3in]{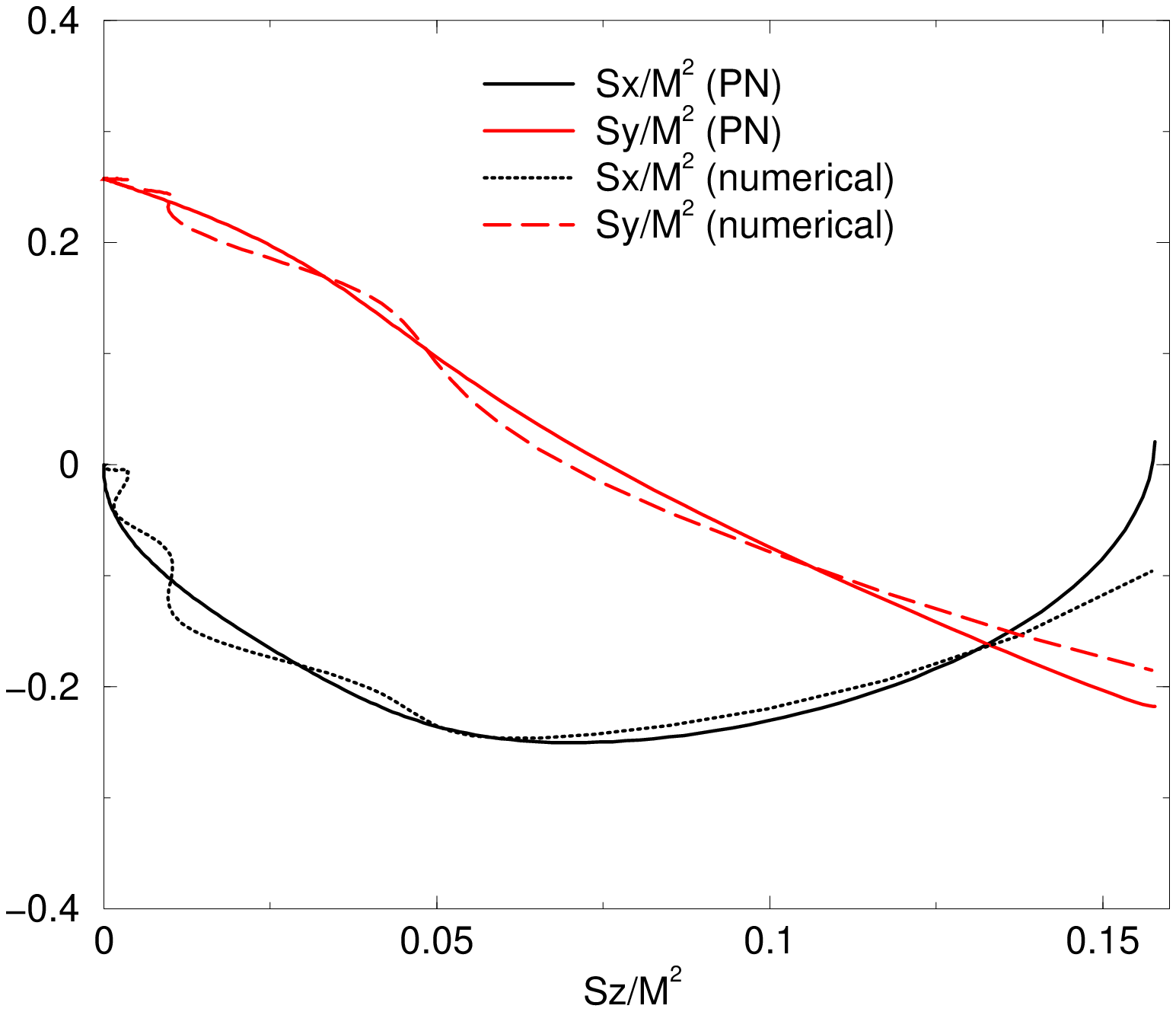}
\caption{ The post Newtonian and numerical $S_x$ and $S_y$ components
of the spin of the individual horizons for the SP3 configuration
as functions of the $S_z$ component of the spin (which
increases monotonically in time prior to the merger).
Note the very reasonable agreement for most of
the evolution. The plot terminates at the merger.}
\label{fig:pn_num_compare}
\end{center}
\end{figure}

In Fig.~\ref{fig:pn_num_compare} we plot the $x$ and $y$ components of
the spin as a function of the $z$ component for both the
Post-Newtonian predicted spins (using numerical tracks) 
and the numerically determined spins.
Plotting the data in this manner removes the ambiguity of assigning
the appropriate Post-Newtonian time to the numerical time coordinate
on the horizon. The qualitative behavior of the spin in our numerical
simulation is consistent with the post Newtonian spin for most of the
evolution (smaller values of $S_z$).

\begin{figure}
\begin{center}
\includegraphics[width=3.3in]{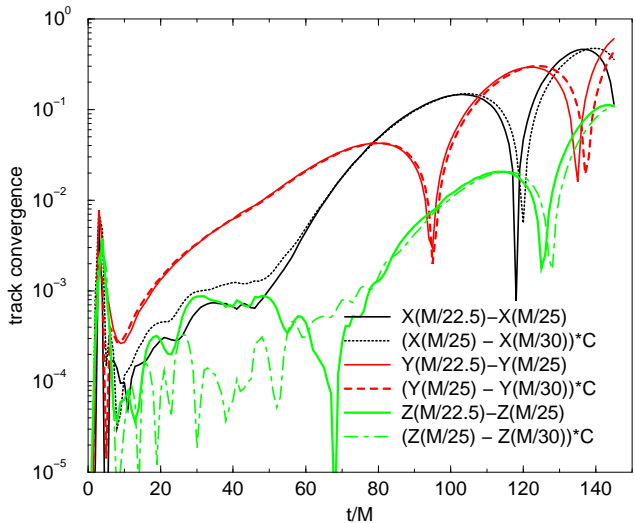}
\caption{The difference in track locations between the $M/22.5$ and
$M/25$ runs as well as the difference in track locations between the
$M/25$ and $M/30$ runs. The latter difference has been rescaled by 
$C=0.88$ to demonstrate third-order convergence. The reduced
order of convergence for $X$ between $t=10M$ and $t=60M$ is likely due
to the coarseness of the grid.}
\label{fig:track_conv}
\end{center}
\end{figure}

The puncture trajectories are third-order convergent as is demonstrated in
Fig.~\ref{fig:track_conv}. The $x$-component of the track appears to
show poorer convergence between $t=10M$ and $t=60M$ but this is likely
due to the coarseness of the grid. Note that the trajectories are
calculated by integrating $\partial_t x^a = -
\beta^a$~\cite{Campanelli:2005dd} at the location
of the punctures. The curve is expected to converge to lower order
because the shift is not smooth on the puncture
(see~\cite{Hannam:2006vv} for a discussion on the behavior of the
evolved fields at the punctures). However, as is shown below, the
constraint violations also show third-order convergence,
which indicates that lower-order convergent effects leak out of the
puncture. These lower-order errors are likely not observed in the
waveform (see below) because of a larger fourth-order error term
dominating the third-order error terms at these resolutions.

The final remnant horizon for the SP3 configuration has mass
 $M_{\cal H}/M = 0.9613\pm0.0007$
with specific spin $S/M_{\cal H}^2 = 0.7215\pm 0.0003$. The remnant spin
components calculated from the approximate Killing vector are
$\vec S_{\rm IH}/M^2 = (-0.045\pm0.001, 0.199\pm0.003, 0.638\pm0.003)$ with
magnitude $S_{\rm IH}/M^2 = 0.669\pm 0.001$. The total
ADM mass and angular momentum (i.e.\ initial mass and angular
momentum) of the system are $M_{\rm ADM}/M = 1.00000\pm0.00005$ and
$\vec J_{\rm ADM}/M^2 = (0, 0.257450, 0.875352)$. Hence $(3.87\pm0.07)\%$ of the mass
and $(23.6\pm0.1)\%$ of the angular momentum were converted into
radiation, with $\delta J_z/M^2 = (0.237\pm0.003)$. The system gained net angular momentum in the $x$-direction
but lost $(22.7\pm0.4)\%$ of its angular momentum in the $y$-direction
 and $(27.1\pm0.3)\%$ of its angular momentum in the $z$-direction.
Thus the binary preferentially radiated angular momentum in the direction
of the initial orbital angular momentum. We also obtained estimates of the
radiated mass and $z$-component of the angular momentum from $\psi_4$
of $(3.8\pm0.1)\%$ and $(0.24\pm0.02)M^2$ respectively, in excelent
agreement with the results from the remnant horizon parameters.
(The errors in the radiated mass and angular momentum from the
waveform are relatively large due to boundary reflection contaminating
the late-time waveform.)
Note that the excellent agreement in $\delta
J_z$ between the final horizon direction measurement and the radiated
$z$-component of the angular momentum indicates that we obtain the
final horizon spin magnitude and direction to within the expect
$2^\circ$ accuracy.
For comparison we also give the remnant spin direction calculated
using the coordinate rotation vectors $\vec S_{\rm coord}/M^2 =
(-0.033\pm0.002, 0.190\pm0.001, 0.6395\pm0.0003)$  (where the errors
are a measure of the flatness of the components of $\vec S_{\rm
coord}$ versus time). The corresponding spin magnitude
$S_{\rm coord}/M^2 = 0.668\pm0.002$ agrees with the norm $S_{\rm IH}$ and
the directions agree to within $1.3^\circ$ (the expected error in the
direction determination is $\sim 2^\circ$).

\begin{figure}
\begin{center}
\includegraphics[width=4in]{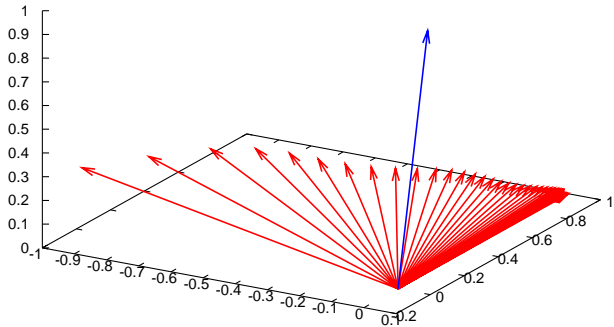}
\caption{The spin-direction of the individual horizons every 
$4M$ during the
spin-precession phase and the final horizon spin-direction for the
SP3 configuration. 
The arrows indicate the spin-direction only, not the magnitude. 
Note the continuous change in the spin-direction during the precession
stage and the discontinuous jump (or flip) to the remnant spin-direction.}
\label{fig:sp3_spin_flip}
\end{center}
\end{figure}

We measure the angle of the spin-flip
both with respect to the initial individual horizon spins and the
individual horizon spins at the merger.
For the SP3 configurations these angles are $72^\circ$
and $71^\circ$ respectively.
In Fig.~\ref{fig:sp3_spin_flip} we show the spin direction of the individual
horizons and spin direction of the remnant horizon. The smooth precession
and discontinuous flip are apparent. Note that the spin flip, unlike
the spin precession, cannot be modeled accurately by a post Newtonian
expansion due to the highly non-linear merger process that converts
roughly $25\%$ of the initial total angular momentum into
gravitational radiation.
The final spin direction for the SP3 configuration is rotated by
$4.0^\circ$ with respect to the initial ADM angular momentum. This
rotation, though small, is larger than the expected error in our
spin-direction algorithm, indicating that there is a small net change
in the direction of the angular momentum.

\begin{figure}
\begin{center}
\includegraphics[width=3.3in]{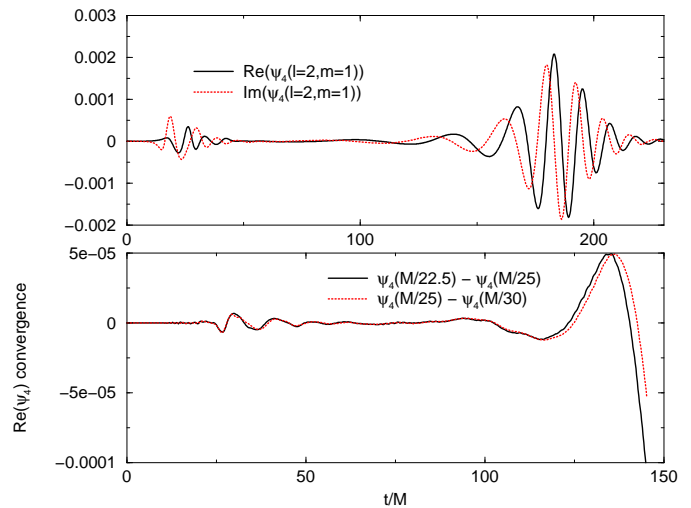}
\caption{The real and imaginary parts of the $(\ell=2,m=1)$ mode of
$\psi_4$ for the SP3 configuration with resolution $h=M/25$ (upper
panel), as well as
a convergence plot of this waveform (lower panel). Note that the
waveform is fourth-order convergent (as is evident by the agreement
of the two differences).}
\label{fig:psi4_2_1}
\end{center}
\end{figure}

The spin direction and puncture trajectories are coordinate dependent
measures of precession. The waveform, on the other hand, should
provide a coordinate independent measure of the precession. To show
the effect of precession on the waveform we examine the $(\ell=2,m=1)$
mode. This mode vanishes identically for the zero-spin, aligned-spin,
and anti-aligned spin cases previously
studied (i.e.\ S0, SC, S++, S-\,-)~\cite{Campanelli:2005dd,Campanelli:2006gf,Campanelli:2006uy,Campanelli:2006fg}.
In Fig.~\ref{fig:psi4_2_1} we show the $(\ell=2,m=1)$ mode at an
extraction radius of $r=15M$ as well as a convergence plot of this
mode (showing fourth-order convergence). The contributions of the
$(\ell=2,m=\pm1)$ modes to the radiated mass are
smaller than the contributions of the dominant $(\ell=2,m=\pm2)$ modes
by a factor of $\sim20$, while the contributions of the
$(\ell=2,m=\pm1)$ modes to the radiated angular momentum are smaller
than the contributions of the dominant $(\ell=2,m=\pm2)$ modes
by a factor of $\sim80$.

\begin{figure}
\begin{center}
\includegraphics[width=3.3in]{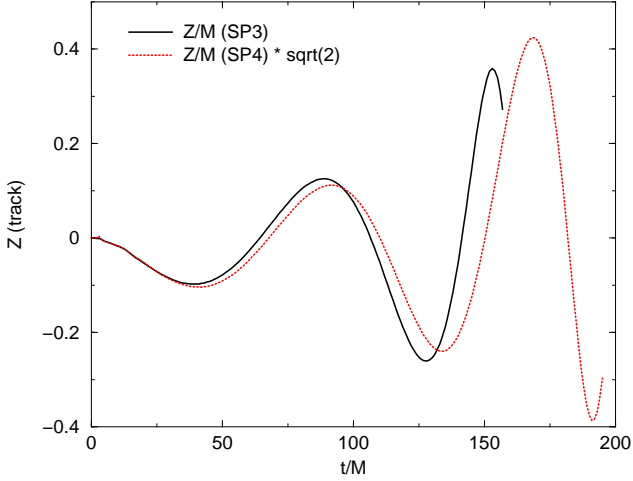}
\caption{The $z$-component of the puncture trajectory (for the puncture
initially located at $x>0$) versus
time for the SP3 and SP4 configurations with resolution $h=M/25$. Note that after rescaling by
$\sqrt{2}$ the two trajectories agree at early times. At later times
the spin-orbit coupling in the partially aligned case delays the
merger; causing the two trajectories to diverge. Both curves terminate
at their respective merger times.}
\label{fig:ztrack_v_config}
\end{center}
\end{figure}

We next examine how the results change when we
set the initial spins closer to the $z$-axis. The SP4 configuration has
the same total spin as the SP3 configuration but at an angle of
$45^\circ$ with respect to the orbital plane. This rotation of the
spin has two significant effects. First, the SP4 configuration has a
significant spin in the same direction as the orbital angular
momentum, and from our previous results~\cite{Campanelli:2006uy,
Campanelli:2006fg} we expect that the binary merger will be delayed
due to the resulting spin-orbit repulsive effect. Second, the amplitude of the
orbital plane precession will be reduced (i.e.\ there is no orbital
plane precession if the spins are
rotated $90^\circ$ with respect to the orbital plane, and the amount of
precession should vary smoothly with angle). In
Fig.~\ref{fig:ztrack_v_config} we show the $z$-component of the first
puncture trajectory versus time for SP3 and SP4, where the latter has
been rescaled by $\sqrt{2}$. Note that at early times the rescaled
tracks agree perfectly. Thus, for a given magnitude of the spin,
 the orbital plane precession has a
$\sin\vartheta$ dependence, where $\vartheta$ is the angle between the
spin and orbital angular momentum. At later times, the spin-orbit
coupling induced delay in the merger becomes evident and the two
tracks no longer agree.

\begin{figure}
\begin{center}
\includegraphics[width=3.3in]{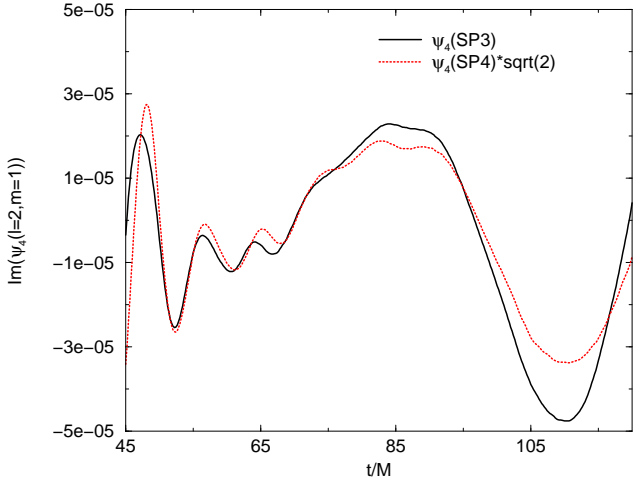}
\caption{The imaginary part of the $(\ell=2,m=1)$ component of
$\psi_4$ for the SP3 and SP4 configurations extracted at $r=15M$ with
a central resolution of $M/25$. Note that between $t=50$ and $t=75$
the two waveforms scale with $\sin\vartheta$ as is evident by the good
agreement between the two waveform after rescaling the SP4 waveform by
$\sqrt{2}$ (i.e.\ $1/\sin(\pi/4)$). However, this scaling breaks down
at later times ($t>85 M$) due to the differences in orbital decay
arising from the increased stability of the SP4 configuration. This
scaling also breaks down at early times because of the non-physical
initial data radiation pulse.}
\label{fig:psi4_2_1_compare}
\end{center}
\end{figure}

Comparing the waveforms from the SP3 and SP4 configurations is
complicated by the fact that the initial data burst masks the
early-time behavior. Nevertheless, there is a small region just after
the initial pulse leaves the system (see
Fig.~\ref{fig:psi4_2_1_compare}) where it is evident that the
$(\ell=2, m=1)$ mode scales with $\sin \vartheta$. However, at later
times the differences in the orbital dynamics destroys this scaling. 

\begin{figure}
\begin{center}
\includegraphics[width=3.3in]{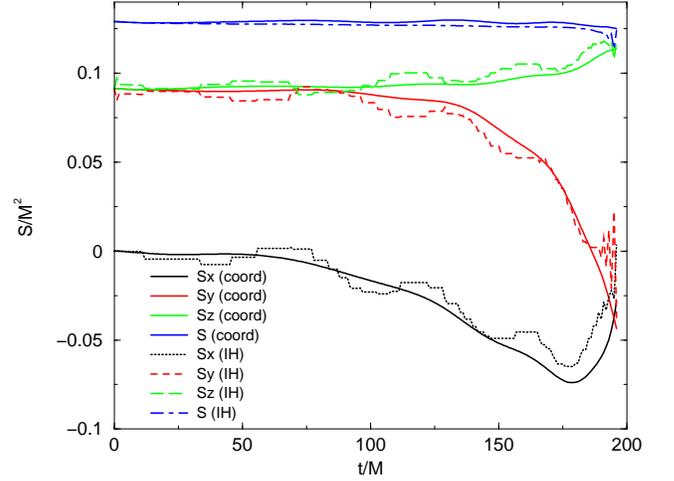}
\caption{The spin components and magnitude of the individual horizons
(the two horizons have equal spins) for the SP4 configuration up to
merger (with resolution $h=M/25$). Note that in this configuration the
spins rotate by $135^\circ$ in the $xy$ plane prior to merger. Also
note that the significant $z$-direction spin-up is once again caused by
a rotation further out of the $xy$ plane. Once again the
Killing vector determination for the spin direction oscillates about the
coordinate definition for the spin direction. The approximate Killing vector
calculation begins breaks down at around $t=180M$.}
\label{fig:sp4_spin_dir}
\end{center}
\end{figure}

\begin{figure}
\begin{center}
\includegraphics[width=3.3in]{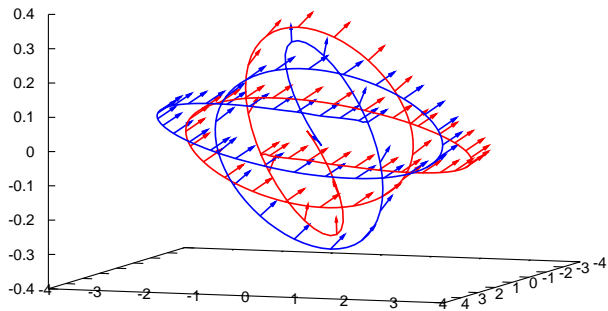}
\caption{The puncture trajectories and horizon spin direction (shown
every $4M$ until merger) for the SP4 configuration. Note that the
$z$-scale is $1/10$th that of the $x$ and $y$ scales.}
\label{fig:sp4_3d}
\end{center}
\end{figure}

\begin{figure}
\begin{center}
\includegraphics[width=3.3in]{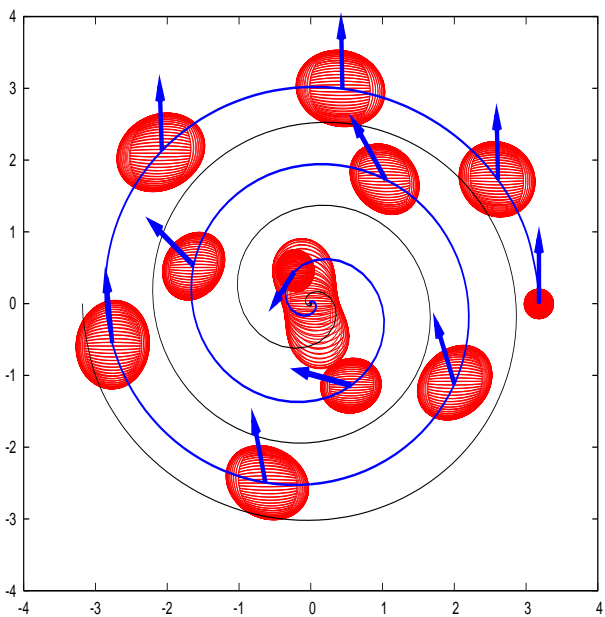}
\caption{The $xy$ plane projection of the orbital track, apparent horizons, and spin 
direction for the SP4 configuration up to merger. The horizons are
given at $t=0, 20M, 40M, ..., 180M, 196M$. The common horizon formed
at $t=195.4M$. The projected spin vector decreases in magnitude at
late time due to the spin rotating further out of the $xy$ plane.
Note that the spin direction precesses by $135^\circ$ in the
$xy$ plane during the merger. The spin of the second black hole
(not shown) is equal to the first. }
\label{fig:sp4_track_spin_xy}
\end{center}
\end{figure}

In
Figures~\ref{fig:sp4_spin_dir},~\ref{fig:sp4_3d},~\ref{fig:sp4_track_spin_xy}
we show the spin magnitude and direction for the SP4 configuration as
well the 3D puncture trajectories and spins and a projection of the
trajectories and spin direction onto the $xy$ plane. Due to the
increased stability of aligned spin binaries, this configuration
completes $2\frac{3}{8}$ orbits prior to merger (compared to
$1\frac{3}{4}$ for SP3).  Consequently spin-precession rotates the
spin vector by an additional $45^\circ$ compared to the SP3
configuration. In this case the Killing vector based calculation of
the spin remains
accurate long enough that the spin-precession rotation beyond 
$90^\circ$ (i.e.\ the local minimum in $S_x$ is observed in $\vec S_{\rm IH}$ 
as well as $\vec S_{\rm coord}$). Note that, once again, there is a significant
spin-up in the $z$-direction caused by a rotation of the spin vector
towards the $z$-axis (rather than a net increase in the spin
amplitude), and  that there is no discernible correlation between the projected
horizon orientation (i.e.\ the orientation of the semi-major axis) and
the projected spin direction.
The individual horizon spins at the merger are $\vec S_{\rm coord} =
(-0.033\pm0.005, -0.041\pm0.003, 0.114\pm0.001)$. Hence 
the total precession angle for the SP3 configuration is $\Theta_p =151^\circ$.

The remnant horizon for the SP4 run formed just as the boundary
reflections began to contaminate the interior. Consequently the error
bounds for the mass and spin of the remnant are higher for SP4 than
SP3. The final remnant mass is $M_{\cal H} = (0.9524\pm0.0002)M$ with
a spin parameter of $S/M_{\cal H}^2 = 0.805\pm0.002$. The spin
components are $\vec S_{\rm coord}/M^2 = (-0.020 \pm 0.003, 0.121 \pm
0.002, 0.720\pm0.002)$, where we used the coordinate-base definition
to calculate $\vec S$. (The Killing vector based calculation of the
spin could not be obtained accurately because the system became
approximately axisymmetric after the boundary errors affected the
remnant spin parameters.) The initial ADM mass and angular momentum
for this system were $M_{\rm ADM}/M = 1.00000\pm0.00005$ and
$\vec J_{\rm ADM}/M^2 = (0, 0.182396, 1.01833)$. Hence,
$(4.76\pm0.02)\%$ of the mass and $(29.4\pm0.2)\%$ of the angular
momentum were converted into gravitational radiation. The system gained
net angular momentum in the $-x$
direction, while losing $(34\pm1)\%$ and $(29.3\pm0.2)\%$ of its
angular momentum in the $y$ and $z$ directions respectively. 
It thus appears that this configuration preferentially radiates
angular momentum in the orbital plane. However, we caution the reader
that the errors quoted for the SP4 configuration for the mass and spin 
do not take into account either possible boundary effects or
finite-difference truncation errors. 
The effect of the radiation on the angular momentum direction is
small, with the SP4 spin direction rotated by only $1.5^\circ$ with respect to
the initial ADM angular momentum.

\begin{figure}
\begin{center}
\includegraphics[width=4in]{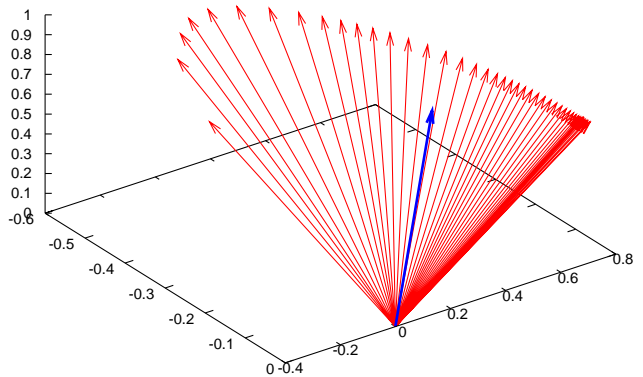}
\caption{The spin-direction of the individual horizons every 
$4M$ during the
spin-precession phase and the final horizon spin-direction for the
SP4 configuration. 
The arrows indicate the spin-direction only, not the magnitude. 
Note the continuous change in the spin-direction during the precession
stage and the discontinuous jump (or flip) to the remnant spin-direction.}
\label{fig:sp4_spin_flip}
\end{center}
\end{figure}

The spin-flip angle of the remnant spin with respect to the initial
individual spins
and the individual spins at merger are $35^\circ$ and $32^\circ$ respectively.
In Fig.~\ref{fig:sp4_spin_flip} we show the spin-direction of the individual
horizons and spin-direction of the remnant horizon. The smooth precession
and discontinuous flip are apparent. Note that in this case boundary
reflections contaminate the waveform at large $r$ prior to the merger.
Consequently we do not obtain reliable measurements for the radiated
mass and angular momentum from the waveform.

\begin{figure}
\begin{center}
\includegraphics[width=3.3in]{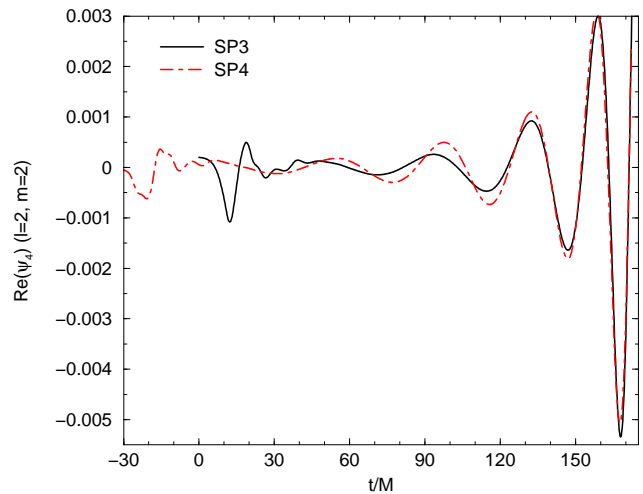}
\caption{The real part of the pre-merger $(\ell=2, m=2)$ component of $\psi_4$
extracted at $r=10M$. The SP4 waveform has been translated by $37M$
and multiplied by a constant phase factor prior to taking the real
part. The plunge part of the waveform begins roughly at $t=168M$. Thus
there are 3.5 cycles of orbital radiation prior to the plunge (but
after the initial data pulse) for SP3
and 4.5 for SP4. This number of cycles match the
number of orbits in
Figs.~\ref{fig:sp3_track_spin_xy}~and~\ref{fig:sp4_track_spin_xy}.
The initial data pulse appears to mask an additional 1/2 cycle of orbital
motion.}
\label{fig:sp3_sp4_wave_2_2}
\end{center}
\end{figure}

As was mentioned above, the calculation of the spin direction is
coordinate dependent. Nevertheless, these particular coordinates show remarkable
agreement between the puncture trajectories and the waveform. In
Fig.~\ref{fig:sp3_sp4_wave_2_2} we show the orbital part of the
$(\ell=2,m=2)$ component of $\psi_4$ extracted at $r=10M$, where
we translated the SP4 waveform and multiplied by a
constant phase in such a way that plunge part of the waveforms
agree~\cite{Baker:2006yw}. Note that the SP3 configuration
shows $\sim4$ cycles of orbital motion prior to the plunge (i.e.\ the last
trough in the plot), while the SP4 configuration shows $\sim 5$ cycles. Thus
we expect that the SP4 track should contain approximately one-half of an
orbit more than the SP3 configuration. Although there is an
uncertainty in the exact location of the beginning of the `plunge'
waveform in the figure, its approximate location will be given by the
formation time of the common horizon at this resolution plus the
coordinate distance to the extraction sphere 
(here we identify the start of the plunge with the first
trough located at $t= 168 = T_{CAH} + 10M$),
the number of cycles in both configurations prior
to the last peak shown is consistent with 1/4 of an orbit more than
the number of orbits (i.e.\ 1.75 and 2.25 respectively) observed in
Figs.~\ref{fig:sp3_track_spin_xy}~and~\ref{fig:sp4_track_spin_xy}.
Interestingly, the number of cycles after the initial pulse of
radiation (3.5 and 4.5 respectively) is in excellent agreement with the
number of orbits observed in the puncture trajectories. Thus
these coordinates appear to reasonably reproduce the orbital dynamics of the
binary. This fidelity by which the coordinates reproduce the merger
dynamics, and the relatively good agreement between the two
measurements of the spin direction, is the motivation for using these
coordinate dependent measurements to measure the spin direction.
In addition, the very good agreement for the radiated $z$-component of
the angular momentum based on the difference between the remnant
 $\vec S_{IH}$ and $\vec J_{\rm ADM}$ for SP3 and the waveform-based
calculation, indicates that this method provides an accurate measurement
of the remnant spin direction and magnitude. We expect the method to
provide more accurate results for the remnant horizon than the
individual horizon because the remnant spacetime is
axisymmetric, and, as pointed out above, there is a natural way to
assign a spin direction to horizons when the spacetime is
axisymmetric. 

\begin{figure}
\begin{center}
\includegraphics[width=3.3in]{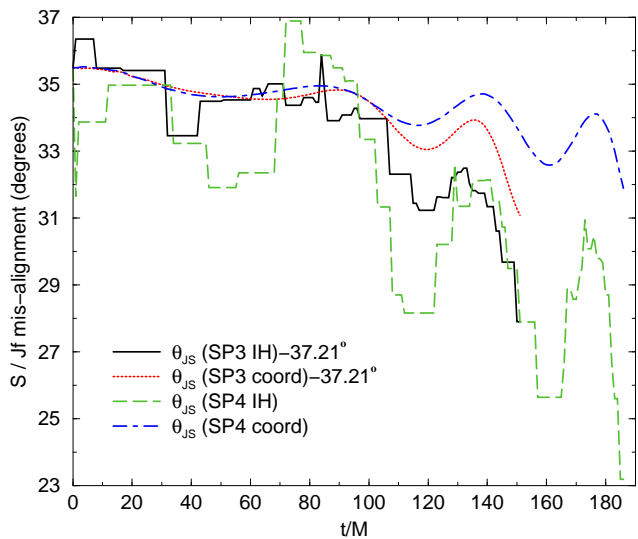}
\caption{The angle in degrees between the individual apparent horizon
spins and the final total angular momentum for the SP3 and SP4
configuration. The SP3 angles have been translated by $-37.21^\circ$
so that the translated SP3 and SP4 angles agree at $t=0$. The curves
terminate at the point when the Isolated Horizon algorithm becomes
inaccurate.}
\label{fig:S_J_angle}
\end{center}
\end{figure}

There appears to be a small trend towards spin--angular momentum
alignment. To observe this effect we measure the angle between the spin
and final angular momentum (i.e.\ the remnant spin),
$\theta_{JS} = \arccos(\hat S\cdot\hat J_{\rm f})$.
 In Fig.~\ref{fig:S_J_angle} we show $\theta_{JS}$ versus
time for the SP3 and SP4 configurations using both $\vec S_{\rm IH}$
and $\vec S_{\rm coord}$.
Both of these measurements show increasingly larger oscillations in
$\theta_{JS}$ versus time. 
The $\vec S_{\rm IH}$ based measurements indicate that the spins in the SP3
configuration migrate $\sim 6^\circ$ towards $\hat J_{\rm f}$, while
the spins in the SP4 configuration migrate $\sim 8^\circ$ towards
$\hat J_{\rm f}$ (we measure the angle at the approximate midpoint
of the oscillation). However, the  $\vec S_{\rm coord}$ based
measurements indicate that the spins in the SP3
configuration migrate by only $\sim 3^\circ$, while the spins in the
SP4 configuration migrate by only $\sim 2.5^\circ$.
It is unclear which measurement is superior. On the one hand $\vec S_{\rm
coord}$ is more strongly coordinate dependent, but, on the other
hand, the horizons become more distorted as they approach each other,
which increases the error in $\vec S_{\rm IH}$ (i.e.\ the horizons
deviate increasingly strongly from axisymmetry). This is an
interesting effect which we plan to study in much more detail with
improved techniques to measure the spin direction.
We summarize the main results from spin-direction calculations 
for the SP3 and SP4 simulation in Table~\ref{table:spinres}.

\begin{table}
\caption{The angle between the initial spin direction and orbital
angular momentum $\vartheta$, the total spin precession angle
$\Theta_p$, spin flip angle
between the initial spin direction and remnant spin direction
$\Theta_{\rm flip}$, and the net change in the $z$-component of the angular
momentum as calculated using the Isolated Horizon spin direction of
the remnant $\delta J_{z}$(IH) and waveform $\delta J_{z} (\psi_4)$ for
the SP3 and SP4 configurations.}
\begin{ruledtabular}
\begin{tabular}{llllll}\label{table:spinres}
Config & $\vartheta$ & $\Theta_p$ & $\Theta_{\rm flip}$  &
$\delta J_{z}$(IH) & $\delta J_{z} (\psi_4)$ \\
\hline
SP3 & $90^\circ$ & $98^\circ\pm2^\circ$ & $72^\circ\pm2^\circ$ & $0.237\pm0.003$ & $0.24\pm0.02$\\
SP4 & $45^\circ$ & $151^\circ\pm2^\circ$ & $35^\circ\pm2^\circ$  & --- & ---\\
\end{tabular}
\end{ruledtabular}
\end{table}

The stability of the spinning binaries is strongly dependent on the
direction of the spin. In Table~\ref{table:tcah} we show the merger
times ($T_{CAH}$) of the SP3 configuration versus resolution and an
extrapolation to infinite resolution, as well as the merger time for
the single SP4 run. The `extrapolated' value of the SP4 merger time
was computed by adding the difference between the extrapolated and
$h=M/25$ merger times for the SP3 configuration to the $h=M/25$ merger
time for the SP4 configuration. The extrapolated values of $176\pm3$
and $\sim214$, for SP3 and SP4 respectively, are in large part
consistent with the results from the aligned-spin binaries if we
replace ${\cal S}$ in Eq.~(\ref{fittM}) with $2 S_{z}/m^2$ (${\cal S}$
is the total spin, hence the factor of 2).  The predicted merger times
are $T_{CAH} = 172\pm1$ and $T_{CAH}=200\pm2$ for the SP3 and SP4
configurations respectively.  The differences between these
predictions and the actual extrapolated merger times can be explained
by the net rotation of the component spins towards the $z$-axis, which
helps stabilize the binaries.

\begin{table}
\caption{Merger times for the SP3 and SP4 configuration versus
resolution, as well as an extrapolation to infinite resolution
(see text for an explanation of the `extrapolation' of the SP4
result.)
}
\begin{ruledtabular}
\begin{tabular}{lll}\label{table:tcah}
Resolution & SP3 & SP4 \\
\hline
$M/22.5$ & $152.0\pm0.2$ & $----$\\
$M/25$ & $157.4\pm0.2$ & $195.4\pm0.2$\\
$M/30$ & $164.2\pm0.2$ & $----$\\
\hline
$M/\infty$ & $176\pm3$ & $\sim214$\\
\end{tabular}
\end{ruledtabular}
\end{table}

\begin{figure}
\begin{center}
\includegraphics[width=3.3in]{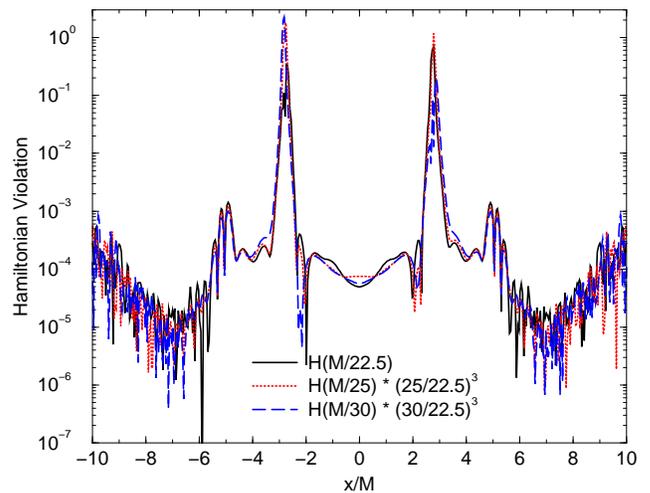}
\caption{The convergence of the Hamiltonian constraint violation for
the SP3 configuration along
the $x$-axis at $t=76M$ when the punctures cross the $x$-axis
for the second time. The Hamiltonian constraint shows third-order
convergence. Points inside the domain of dependence of the boundary
have been excluded. The high-frequency features near the outer boundary
are due to the extreme fisheye de-resolution and converge with
resolution.}
\label{fig:sp3_hc_conv}
\end{center}
\end{figure}

\begin{figure}
\begin{center}
\includegraphics[width=3.3in]{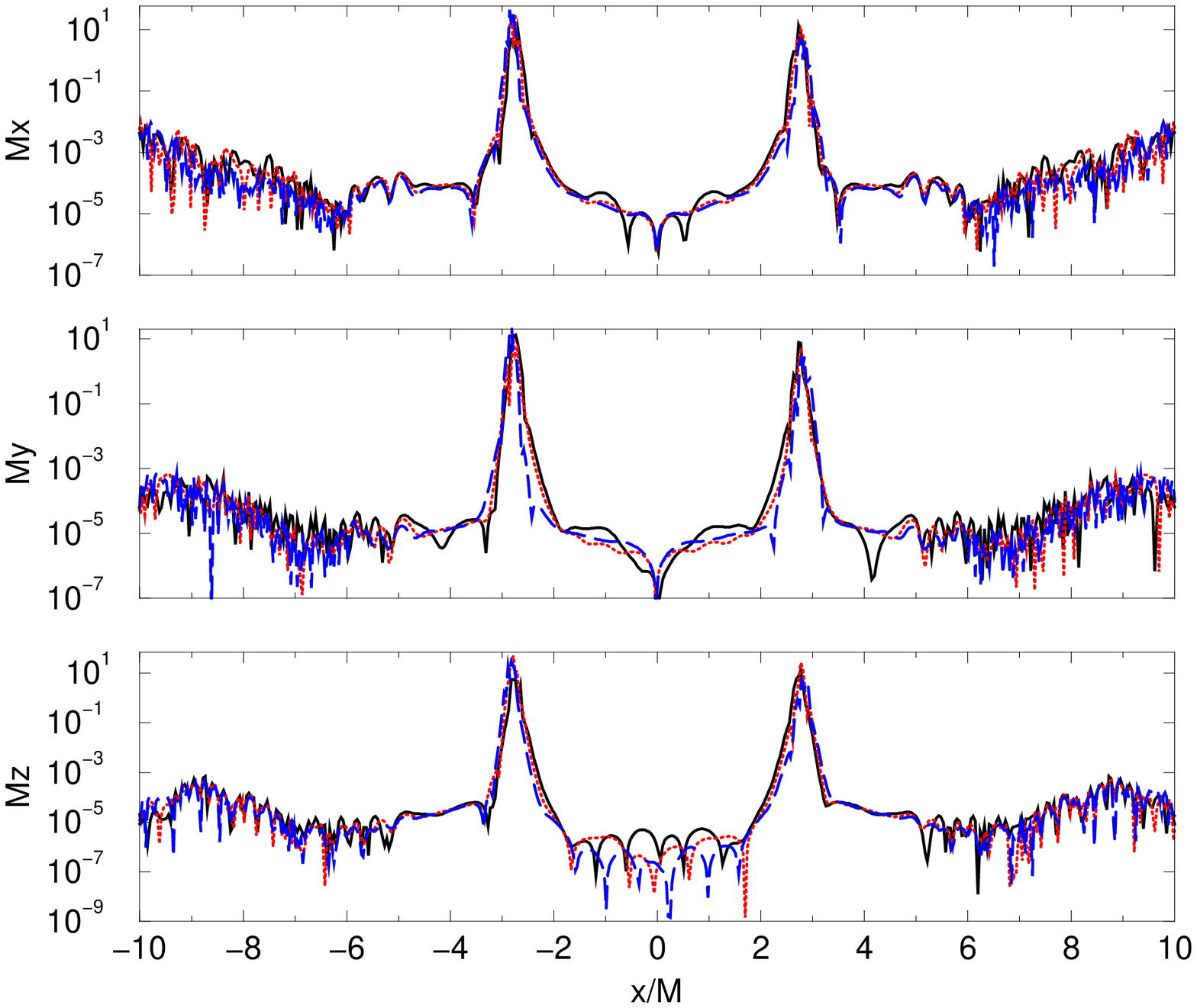}
\caption{The convergence of the Momentum constraint violation for
the SP3 configuration along
the $x$-axis at $t=76M$ when the punctures cross the $x$-axis
for the second time. The Momentum constraint shows third-order
convergence. Points inside the domain of dependence of the boundary
have been excluded. In each panel the solid (black) curve, dotted (red) curve
and dashed (blue) curves are the $M/22.5$, $M/25$ and $M/30$
constraint violations respectively. The constraints have been rescaled
by $(h_l/h)^3$ ($h_l=M/22.5$). The high-frequency features near the
outer boundary are due to the extreme fisheye de-resolution and
converge with resolution.}
\label{fig:sp3_mc_conv}
\end{center}
\end{figure}

\begin{figure}
\begin{center}
\includegraphics[width=3.3in]{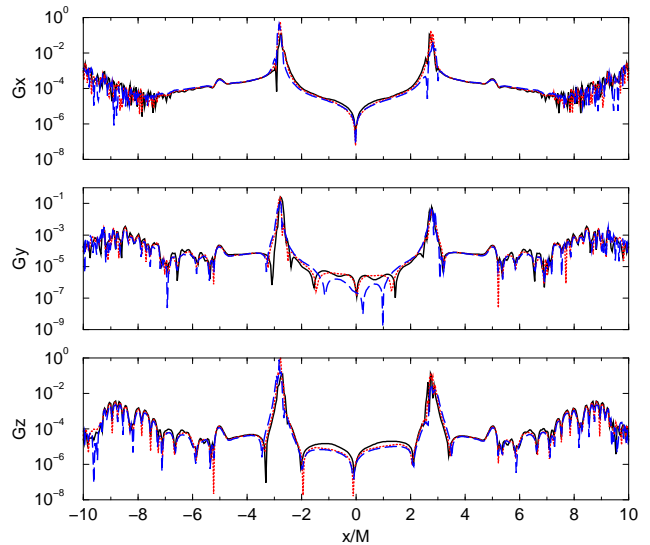}
\caption{The convergence of the BSSN constraint violation for
the SP3 configuration along
the $x$-axis at $t=76M$ when the punctures cross the $x$-axis
for the second time. The BSSN constraint shows third-order
convergence. Points inside the domain of dependence of the boundary
have been excluded. In each panel the solid (black) curve, dotted (red) curve
and dashed (blue) curves are the $M/22.5$, $M/25$ and $M/30$
constraint violations respectively. The constraints have been rescaled
by $(h_l/h)^3$ ($h_l=M/22.5$). The high-frequency features near the
outer boundary are due to the extreme fisheye de-resolution and
converge with resolution.}
\label{fig:sp3_gtc_conv}
\end{center}
\end{figure}

We conclude this section by showing that the  constraint
violations converge to third-order. Although the code uses purely
fourth-order stencils, lower order errors both from the lower
differentiability of the evolved fields at the punctures, as well as
from the second-order accuracy of the initial data, lead to a global
third-order error in the constraint violation.
Figures~\ref{fig:sp3_hc_conv},~\ref{fig:sp3_mc_conv},~\ref{fig:sp3_gtc_conv}
show the Hamiltonian constraint, Momentum constraint, and BSSN
constraint ($G^i = \tilde \Gamma^i + \partial_j \tilde g^{ij}$)
violations at $t=76M$ (the time when the punctures cross the $x$-axis
for the second time) along the $x$-axis for the SP3 configuration.
 The constraint violations have
been multiplied by $(h_l/h)^3$ (where $h_l = M/22.5$) to demonstrate third-order
convergence. Note that the high-frequency features near the outer
boundary are due to the extreme fisheye de-resolution near the outer
boundary and converges to zero with resolution. 

\section{Conclusion}
\label{sec:summary}

In this paper, we evolved systems of equal-mass and equal-spin
black-hole binaries with initial spins aligned perpendicular to, and
$45^\circ$ to, the orbital angular momentum.  We observed the combined
effects of spin and orbital plane precession as predicted by post
Newtonian theory, with dramatic, total precessions of $\sim 98^\circ$
and $\sim 151^\circ$ in the SP3 and SP4 simulations respectively.  Both
configurations resulted in large spin flips between the individual
horizon spin directions and the final remnant direction, with the SP3
configuration showing a spin flip of $\sim 72^\circ$ and the SP4
configuration (which had spins initially more closely aligned with the
orbital angular momentum) resulting in a spin flip of $\sim 34^\circ$.
We see a possible small trend to spin---orbital angular momentum
alignment, but no evidence for significant spin-up of the
black-hole spins.

Although the configurations studied here are parity-symmetric, this
symmetry was only chosen to reduce the memory footprint of the simulations 
(allowing for higher resolution runs); it does not
affect the stability of the `moving punctures' algorithm. Notably, the
punctures move out of the $xy$ plane and can get arbitrarily close to the numerical
gridpoints. Despite this symmetry, the SP3 and SP4 configurations
display most of the significant spin-orbit coupling effects associated
with spinning binaries: spin and orbital plane precession, spin
flips, and enhanced stability of the semi-aligned configuration.
The only significant spin-orbit coupling effect not shown by these
configurations is a spin-orbit induced kick of the remnant hole.
In order to see these kicks we would need to evolve configurations
without parity symmetry.

Our methods for calculating the spin direction
produce reasonable results for our choice of gauge
conditions. Future work will concentrate on improving this calculation
with alternative choices of the gauge parameters (e.g.\ $\eta$ in the
Gamma-driver shift and various different choices of initial values for
the lapse function) and with alternative forms of the Gamma-driver
shift condition. Notably, the lack of agreement between the location
of the horizon semi-minor axis and the spin direction indicates that
these coordinates are not yet ideal. In
addition, it would be useful to calculate independent measures of
the quality of the approximate Killing vector, for example the norm of
the Lie derivative of the 2-metric on the horizon $\mathcal{L}_\varphi
q_{ab}$. It is interesting to note that
the purely coordinate measurement $\vec S_{\rm coord}$ gives reasonable
results for the spin direction and amplitude, and since this
calculation is both more robust (i.e.\ the approximate Killing vector
may not exist) than the approximate Killing vector
calculations and easier to implement, it may prove to be a convenient
measurement of the spin for those codes that have not implemented the
approximate Killing vector finding algorithm.

From the mathematical side, further investigation is required to make
the definition of the spin vector more rigorous and gauge independent.
There are some interesting questions deserving further attention.
Is it possible to meaningfully compare the spin vector defined at the
black hole with the vector defined at spatial infinity?  What about
comparing the spin vector of the final black hole with that for the
individual black holes that we start with?  Does it matter that they
are all calculated on different dynamical horizons?  With regard to
the latter question, it is possible that the dynamical horizons for
the individual black holes are smoothly connected with the final
dynamical horizon through the appearance of marginally trapped
surfaces lying between the common outer horizon and the two inner
horizons.  This scenario is suggested by initial numerical studies
\cite{Schnetter:2006yt}, but further analytical and numerical work is
required to confirm this and to fully understand the dynamics of
marginally trapped surfaces.

We also revisited the question of tidal locking leading to corotation
in close black hole binaries.  Corotation implies both that the
spin directions are aligned with the orbital angular momentum and that
the horizon frequency (the horizon frequency is the angular speed of
locally-non-rotating observers as they pass through the horizon, as
seen by stationary observers at infinity) is equal to the orbital
frequency. We found in a previous study~\cite{Campanelli:2006fg}
that the spin-up of the holes is too small (by
two orders of magnitude) to reach corotation. In this paper we
observed signs of alignment of the spin with the total angular
momentum.  The variations in the direction
observed are a few degrees during the last two orbits. Further
numerical and analytic (higher PN order) studies will be needed to see
if this effect is strong enough during the slow inspiral phase to
drive the binary toward spin--orbit alignment.

Both the final magnitude, and the final direction, of a black-hole
binary's remnant spin are of astrophysical interest: the former
determines the efficiency of gravitational accretion, and the latter
is reflected in the orientation of the inner accretion disk and
(indirectly) in the launching direction of a jet.  Our simulations are
the first to follow the time dependence of the spin orientations in
black hole mergers with initially mis-aligned spins, and the first to
verify the spin-flip phenomenon: the sudden reorientation in spin axis
that takes place when the binary's orbital angular momentum is
converted into spin in the final stages of the
merger~\cite{Merritt:2002hc}. In addition to
influencing the gravitational wave forms, the spin evolution would
also be reflected in any electromagnetic signature due to gas in orbit
around the black holes.  Predicting the latter signature  is beyond
the scope of the present paper but is a fruitful topic for further
study.

\acknowledgments 
 We thank Alessandra Buonanno and Kip Thorne for valuable discussions
 about spin-flips, and Vicky Kalogera for bringing up the papers
 related to the data analysis of precessing binaries. We thank
 Erik Schnetter for technical support and for providing the thorns to
 implement Pi-symmetry boundary conditions, and Marcus Ansorg
 for providing the {\sc TwoPunctures} initial data thorn.  We
 gratefully acknowledge the support of the NASA Center for
 Gravitational Wave Astronomy at University of Texas at Brownsville
 (NAG5-13396) and the NSF for financial support from grant
 PHY-0722315.  Computational resources were provided by the Funes
 cluster at UTB, the Lonestar cluster at TACC, and the Tungsten
 cluster at NCSA.

\appendix
\section{Previous Studies}
\label{sec:review}
In a previous paper we reported the first fully-nonlinear studies of 
highly-spinning black-hole binaries~\cite{Campanelli:2006uy}, where we 
found that the spin can profoundly affect the orbital dynamics of the last
pre-merger stages. In Ref.~\cite{Campanelli:2006uy} we studied cases
where the spins were aligned or counter aligned with the orbital
angular momentum. As a result of the spin-orbit coupling the merger times
dramatically changed with respect to the non-spinning case. For
example, for initial data corresponding to a quasi-circular orbit
with period $T\sim125M$ and orbital frequency $\omega=0.05/M$, the
non-spinning holes would orbit twice before merging into a single
horizon, while the spinning holes aligned with the orbital angular momentum
and spinning at a rate $S/m^2=0.75$ (where $S$ is the magnitude of the
spin angular momentum and $m$ is the mass of the black hole)
 would orbit three times before
merger.  The anti-aligned spinning holes with specific spins
$S/m^2=-0.75$ would only complete one orbit before the common event
horizon formed. These results can be summarized by a linear fit to the
Richardson extrapolated merger times $t_{CAH}$ (formation time of the
first common apparent horizon) of the most accurate runs
with $S/m^2=0.0$, $S/m^2=0.1$, and $S/m^2=-0.757$ (see
 Refs.~\cite{Campanelli:2006fg,Campanelli:2006uy})
\begin{equation}\label{fittM}
\frac{t_{CAH}}{M}=(172\pm1)+(40\pm2)\,{\cal S},
\end{equation}
where ${\cal S}\equiv(S_1/m_1^2+S_2/m_2^2)_I$.
Note that extrapolating to maximally spinning holes gives a merger time
(from orbital $\omega=0.05/M$) of $87M$ and $255M$ for anti-aligned and aligned
spins respectively.

\begin{figure}
\begin{center}
\includegraphics[width=3.3in]{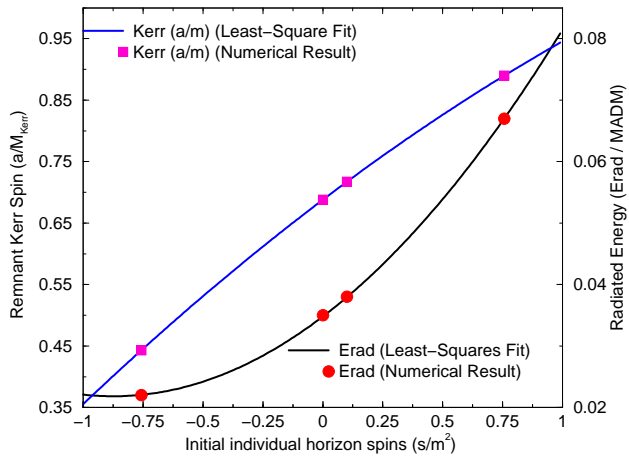}
\caption{A linear-least squares fit of the remnant Kerr spin parameter
(left $y$-axis)
and radiated energy (right $y$-axis) for the merger of equal-mass 
equal-spin binaries with spins pointing along (or in the opposite
direction to) the orbital angular momentum. The fits have the functional
form $y = c_0 + c_1 (S/m^2) + c_2 (S/m^2)^2$.}
\label{fig:spin_extrap}
\end{center}
\end{figure}

Extrapolation to maximally rotating black holes aligned with the
orbital angular momentum leads to remnant black holes having a sub-maximal 
specific rotation parameter $S/m^2<0.95$ which implies one cannot generate
extreme rotating black holes or violate the cosmic censorship hypothesis
starting from orbiting black holes (see Fig.~\ref{fig:spin_extrap}).

A quadratic fit to the remnant black hole of the merger of aligned or
anti-aligned spinning holes  produces
\begin{equation}\label{fitaM}
  (S/M_{H}^2)|_{R} = 0.6879 + 0.1476 \left({\cal S}\right) 
             -0.00935 \left({\cal S}\right)^2,
\end{equation}
while a fit to the energy radiated versus the initial
individual spins yields
\begin{eqnarray}\label{fitEM}
  \frac{E_{rad}}{M} = 0.0348 + 0.01485\left({\cal S}\right) +
   0.00425 \left({\cal S}\right)^2,
\end{eqnarray}

While we expect more simulations of spinning black holes for other
values of the individual spins and with even higher accuracy will
give improved fits, Eqs.~(\ref{fittM})-(\ref{fitEM}) already
provide valuable information for data analysts and for theoreticians
modeling the merger of spinning black-hole binaries with 
post-Newtonian or `Kludge' waveforms~\cite{Buonanno:2005xu}.

This differential orbital dynamics in turn also notably changes 
waveforms (see Figs.~1-3 in Ref.~\cite{Campanelli:2006uy}).

We then explored changes in the magnitude
of the spin due to tidal effects in binaries and the transfer of
orbital angular momentum to spin and {\it vice versa}~\cite{Campanelli:2006fg}.
Those studies concluded that
it is very unlikely that black-hole binaries become tidally locked in a
corotating state during the last orbital stages. We considered two
representative cases, one starting with initially non-spinning black
holes and tracked the spin-up during the last two orbits before
merger. The second simulation began with the binary in an
instantaneously corotating state at the same starting
point and again tracked the spin-up of the individual holes. In both
cases the spin-up was two orders of magnitude smaller than that needed
to lock the binary into a corotating state.

\bibliographystyle{apsrev}
\bibliography{../../../Lazarus/bibtex/references}
\thebibliography{letter}

\end{document}